# Fast camera studies at an electron cyclotron resonance table plasma generator[a)]

R. Rácz,[1,2,b)] S. Biri,[1] P. Hajdu,[2] and J. Pálinkás[2]

[1]*Institute for Nuclear Research (ATOMKI), H-4026 Debrecen, Bem tér 18/c, Hungary*

[2]*Department of Experimental Physics, University of Debrecen, H-4032 Debrecen, Egyetem tér 1, Hungary*

A simple table-size ECR plasma generator operates in the ATOMKI without axial magnetic trap and without any particle extraction tool. Radial plasma confinement is ensured by a NdFeB hexapole. The table-top ECR is a simplified version of the 14 GHz ATOMKI-ECRIS. Plasma diagnostics experiments are planned to be performed at this device before installing the measurement setting at the "big" ECRIS. Recently, the plasma generator has been operated in pulsed RF mode in order to investigate the time evolution of the ECR plasma in two different ways. (1) The visible light radiation emitted by the plasma was investigated by the frames of a fast camera images with 1 ms temporal resolution. Since the visible light photographs are in strong correlation with the two-dimensional spatial distribution of the cold electron components of the plasma it can be important to understand better the transient processes just after the breakdown and just after the glow. (2) The time-resolved ion current on a specially shaped electrode was measured simultaneously in order to compare it with the visible light photographs. The response of the plasma was detected by changing some external setting parameters (gas pressure and microwave power) and was described in this paper.

## I. INTRODUCTION

All over the world, ECR ion sources are operating mainly in continuous wave (CW) mode utilizing high gas efficiency which is achieved in the order of several 10 ms after switching on the microwave heating.

In many cases, however, pulsed operation mode also can be very important. During the pulsed mode operation, various plasma parameters regimes can be observed which are hidden in the CW-mode. Beside the well-known afterglow transient[1] (appearing just after the pulse end) in 2004, a new plasma regime was discovered: transient ion current peak was observed at the very beginning of the gas breakdown.[2] This phenomenon is named preglow. Preglow was observed for low and medium charge state ions. In 2008 experimental evidence and a possible theoretical explanation were given.[3] It was found that at the very beginning of the breakdown the electron energy distribution function (EEDF) is determined by a superadiabatic interaction of the electrons with the microwave. As the plasma density further increases, more power is necessary to sustain such EEDF and finally the absorbed microwave power becomes insufficient. At this moment, the EEDF turns into a Maxwellian. The average energy drops down which results fast plasma outflow from the ECRIS.

Since 2009 many aspects of the preglow peak were investigated in order to precisely test and extend the theory. The plasma breakdown was studied by measuring biased electrode current[4] and plasma potential,[5] and detecting the plasma bremsstrahlung[6] with 100 $\mu$s temporal resolution.

These studies pointed out the importance of the seed electrons of the plasma, and by the X-ray measurements, the evidence of the superadiabatic EEDF was verified. But at same time lower photon energy measurements were suggested in order to collect more information on the other part of the electron component.

The recently published average electron energy simulations[7] show a valley in the time scale just after the superadiabatic stage. Our previous works clearly show that the visible light (VL) irradiation emitted by the plasma is strongly determined by the cold electron component of the electron cloud.[8] Therefore, we are expecting transient behavior in the VL caused by the average energy drop. It was decided to make spatially resolved fast camera movies in order to investigate the time evolution of the emitted VL and to
measure the time-resolved ion current on a specially shaped grounded electrode simultaneously for comparison. Because of the different spatial positions of the different energy electrons, we are expecting different evolution characteristic of the plasma at different spatial positions.

## II. EXPERIMENTAL SETUP

Plasma provided by the compact ECR table plasma generator built in our laboratory was used for this measurement. In Fig. 1, the schematic of the experimental setup is seen. The plasma generator consists of a relatively big plasma chamber (1) (ID = 10 cm, L = 40 cm) in a thin NdFeB hexapole magnet (2) with independent vacuum and gas dosing systems. There is no axial magnetic trap and there is no extraction.

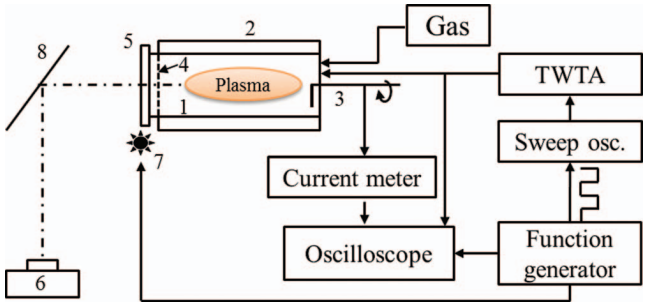

FIG. 1. Schematic of the measurement setup.

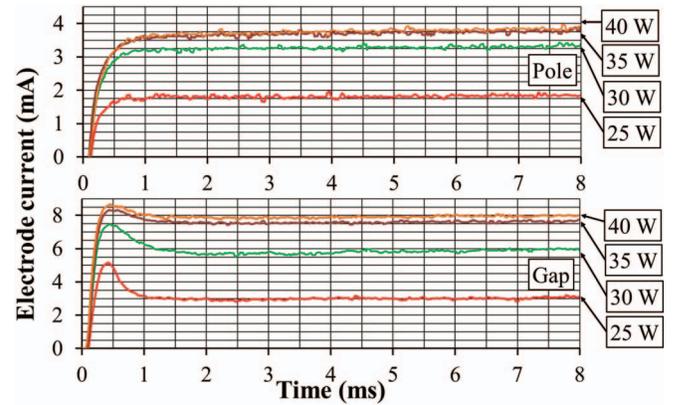

FIG. 2. The first 8 ms of the ion current signal at the pole and at the gap positions as function of microwave power.

For the experiments described in this paper the plasma source operated in pulsed operation mode. For microwave coupling low power (40 W) TWTA was used and the microwave frequency was 9 GHz. The microwave signal was provided by HP 8350B sweep oscillator with HP 8354A plug-in (5.9–12.4 GHz). It was modulated by external rectangular signal provided by HP 8116A function generator. The repetition rate of the pulsed plasma was 10 Hz with 50% duty ratio. The ramp up time of the microwave power was measured by Schottky diode to be on the order of microseconds. The laboratory air was injected to the plasma chamber as working gas.

The ion current was measured as analogue output of the Keithley 6485 type pA-meter using Tektronix TDS 3000 type digital phosphor oscilloscope. The electrode (3) was connected to the ground across the current meter. The rise time of the current signal was 0.15 ms. The movable and rotatable electrode has special oval shape with 15 mm width and with 45 mm length. The electrode can be rotated to many azimuthal positions in order to collect particle coming from representative areas. In order to check the composition of the obtained current voltage-current characteristics were recorded at different setting parameters of the plasma source and at different (azimuthal) positions of the electrode. It proved that the measured current on the grounded electrode is mainly ionic current.

The end of the plasma chamber was closed by a stainless steel mesh (4) (mesh size is 1 mm, wire diameter 0.2 mm) inside the plexiglas window (5) in order to form a closed resonance cavity and to avoid the losses of the microwave power. The VL emission of the plasma was detected by the frames of the movie taken by Casio Exilim FH 25 type fast camera (6) with 1 ms temporal resolution and with $224 \times 64$ pixel spatial resolution. Numerical information was observed through the Analogue to Digital Unit (ADU) values of the pixels of the frames. The frames were triggered by a quick response Light Emitted Diode (7). The LED was switched on when the microwave power was injected to the chamber and the light emission of the LED was also recorded by the camera. A mirror (8) was placed in 45° angle at a distance of 50 cm from the middle plane of the plasma chamber to set the camera at a safe perpendicular 30 cm distance from the mirror and from the axes of the plasma chamber. The frame rate of the movies was set to 1000 frame/s while the shutter speed was always 1/2000 s. The sensitivity of the sensor was set to 100 ISO value in order to record the frames with minimal noise. The Iris value was set to 3.6. The camera settings provided more than 40 cm DOF (Depth of Field) length.

The response of the plasma was detected by changing some external setting parameters. The microwave power ($P_{mw}$) was changed from 25 W up to 40 W and the gas pressure ($P_{gas}$) was changed between $3.5 \times 10^{-4}$ mbar and $1.5 \times 10^{-3}$ mbar. The time evolution of the ion current was measured at two representative positions: magnetic pole position and between two magnetic poles (hereafter it will be named as gap) position. The ADU values of the full-size images and of the selected areas (from pole and gap) were analyzed.

## III. TIME EVOLUTION OF THE ION CURRENT

At first, time evolution of the ion current measured at the pole and at the gap positions as function of microwave power was recorded (Fig. 2).

The families of these curves show the most interesting behavior at the first three milliseconds: transient (preglow) peaks only at the gap position were observed. The saturation current values increase with the microwave power. The rise time of the current curves measured at the pole position is 0.5 ms and is irrespective of the microwave power. After switching off the microwave power the signal disappears within 0.5 ms. The higher the microwave power the lower the preglow peak relative to the saturation current was measured in case of the gap position. Fig. 3 shows the time evolution of the normalized ion current measured at the gap position as function of the gas pressure. The saturation current grows exponentially with the gas pressure. However, the amplitude of the preglow peak relative to the saturation current as function of gas pressure can be fitted by exponentially decay curve. After switching off the microwave power the signal disappears within 0.5 ms.

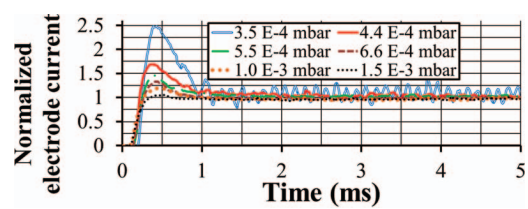

FIG. 3. The first 5 ms of the ion current signal measured at the gap position as function of gas pressure.

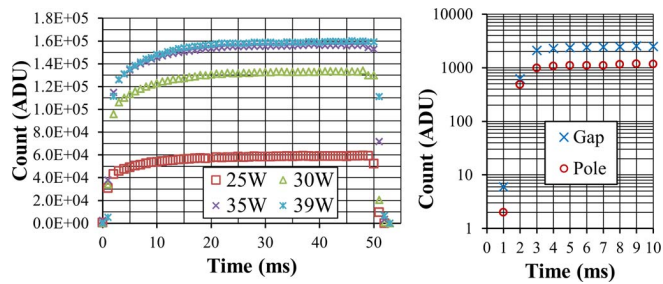

FIG. 4. Time evolution of the visible light photons as function of microwave power (a). First 10 ms ROI counts from pole and gap positions at 30 W (b). ($P_{gas} = 5.5 \times 10^{-4}$ mbar).

## IV. TIME EVOLUTION OF THE VL PHOTONS

Time evolution of the total intensity of the visible light photons emitted by the plasma was measured as function of the microwave power (Fig. 4(a)) and of the gas pressure.

At first the intensity is rapidly increasing, after slightly increasing and finally saturating. After the pulse end it falls down within 1 ms. The saturation intensity successively increased as a function of the microwave power and of the gas pressure and the time structures of the curves are irrespective of them. Preglow peak was not detected.

In the course of the grounded electrode investigation preglow peak appeared at the gap position. Therefore, counts from pole and gap position were collected. The time structures of these curves are similar (Fig. 4(b)). However, the time scale of the transient is (showed by the electrode current) less than 1 ms which is commensurate with the temporal resolution of the fast camera and the direct observation was not expected. In order to investigate the relation of the time evolution of the different spatial part of the plasma in the VL range ratio of the counts coming from gap and pole position was calculated (R(VL)) at each time step. Such ratios do show transient behavior in the first ms in all cases (Fig. 5(a)). For comparison in Fig. 5(b) the amplitudes of the transient current peaks relative to the saturation current values (R(C)) and value of the R(VL) at the first ms were plotted. Strong correlation was found between them. Higher the gas pressure lower the R(C) and the R(VL) in the first ms relation was manifested.

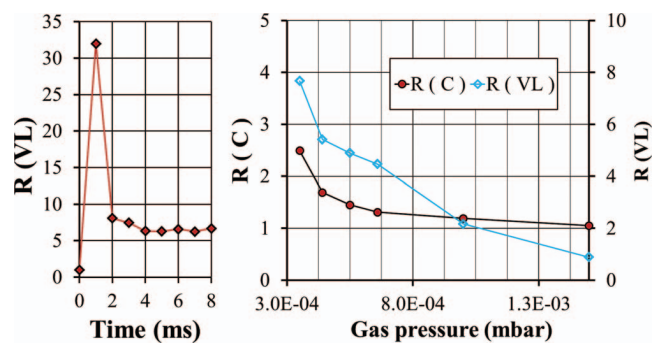

FIG. 5. Typical tendency of the R(VL) values ($P_{mw} = 23W$ and $P_{gas} = 5.5 \times 10^{-4}$ mbar) in the first 8 ms (a). The R(VL) values in the first ms ($P_{mw} = 25W$) and the R(C)s as function of gas pressure (b).

## V. CONCLUSION

While the saturation current on the grounded electrode strongly depends on the gas dose rate and the microwave power, the time structure of the curves are irrespective of it. Transient peak at the gap position was found only. The saturation counts at the pictures are strongly depend on the gas dose rate and microwave power, but the time structures of the curves are irrespective of it. No preglow peak was found because the length of the transient is less than 1 ms. However, calculating the ratio of the counts coming from different spatial position of the plasma the same conclusion can be drawn as at the ion current: the breakdown transient is strongly localized to the gap position. The plasma in visible light range starts at the magnetic gap position.

Our plan for the future is to continue the spatially and time-resolved VL measurements using fast camera with higher temporal resolution and extending the investigation to ECR-traps with B-minimum geometry.

## ACKNOWLEDGMENTS

The research leading to these results has received funding from the European Union Seventh Framework Program FP7/2007–2013 under Grant Agreement No. 262010 – ENSAR. The EC is not liable for any use that can be made on the information contained herein. This work was also partly supported by the TAMOP 4.2.2.A-11/1/KONV-2012–0036 project, which is co-financed by the European Union and European Social Fund.